\documentstyle[12pt,epsfig]{article}

\newlength{\dinwidth}
\newlength{\dinmargin}
\begin{document}

\def\bold#1{\setbox0=\hbox{$#1$}%
     \kern-.025em\copy0\kern-\wd0
     \kern.05em\copy0\kern-\wd0
     \kern-.025em\raise.0433em\box0 }
\def\slash#1{\setbox0=\hbox{$#1$}#1\hskip-\wd0\dimen0=5pt\advance
       \dimen0 by-\ht0\advance\dimen0 by\dp0\lower0.5\dimen0\hbox
         to\wd0{\hss\sl/\/\hss}}
\def\lq{\left [}
\def\rq{\right ]}
\def\II{{\cal I}}
\def\LL{{\cal L}}
\def\VV{{\cal V}}
\def\BB{{\cal B}}
\def\MM{{\cal M}}
\def\pv{\mbox{\bf p}}
\def\ovl{\overline}
\def\pr{^\prime}
\newcommand{\be}{\begin{equation}}
\newcommand{\ee}{\end{equation}}
\newcommand{\bea}{\begin{eqnarray}}
\newcommand{\eea}{\end{eqnarray}}
\newcommand{\ba}{\begin{array}}
\newcommand{\ea}{\end{array}}
\newcommand{\nn}{\nonumber}
\newcommand{\dd}{\displaystyle}
\newcommand{\bra}[1]{\left\langle #1 \right|}
\newcommand{\ket}[1]{\left| #1 \right\rangle}
\newcommand{\spur}[1]{\not\! #1 \,}
\newcommand{\nor}[1]{{}_{\times}^{\times} #1 {}_{\times}^{\times}}

\thispagestyle{empty}
\vspace*{1cm}
\rightline{Napoli DSF-T-44/99}
\vspace*{2cm}
\begin{center}
  \begin{LARGE}
  \begin{bf}
A Conformal Field Theory description of the Paired and
parafermionic states in the Quantum Hall Effect
\footnote{Work supported in part by EC n. FMRX-CT96-0045.}
  \end{bf}
  \end{LARGE}

  \vspace{8mm}

  \begin{large}
Gerardo Cristofano ~~~~ Giuseppe Maiella \\ and \\ Vincenzo Marotta
  \end{large}
  \vspace{1cm}

\begin{it}
Dipartimento di Scienze Fisiche  \\
 Universit\'{a} di Napoli ``Federico II'' \\
and \\ INFN, Sezione di Napoli \\ Mostra d'Oltremare
Pad.19-I-80125 Napoli, Italy\footnote{E:mail:
gerardo.cristofano(giuseppe.maiella;vincenzo.marotta)@na.infn.it}
\end{it}
\end{center}
\begin{quotation}

\begin{center}
\begin{bf}
Abstract\\
\end{bf}\end{center}
We extend the construction of the effective conformal field
theory for the Jain hierarchical fillings proposed in \cite{CGM}
to the description of a quantum Hall fluid at non standard
fillings $\nu =\frac{m}{pm+2}$. The chiral primary fields are
found by using a procedure which induces twisted boundary
conditions on the $m$ scalar fields; they appear as composite
operators of a charged and neutral component. The neutral modes
describe parafermions and contribute to the ground state wave
function with a generalized Pfaffian term. Correlators of $N_e$
electrons in the presence of quasi-hole excitations are
explicitly given for $m=2$.
\noindent

$\vspace{1cm}$


\end{quotation}

\newpage
\baselineskip=18pt
\setcounter{page}{2}

The experimental evidence of a Hall plateau at filling
$\nu=\frac{5}{2}$ has recently spurred a renewed interest in a
deeper understanding of the underlying physics at plateaus which
do not fall into the hierarchical scheme \cite{jain}. To such an
extent a pairing picture, in which pairs of spinless or
spin-polarized fermions condense, has been presented \cite{MR}
for the non-standard fillings $\nu=\frac{1}{q}$, $q>0$ and even.
As a result the ground state gets described in terms of the
Pfaffian (the so called Pfaffian state) and the non-Abelian
statistics of the fractional charged excitations evidenced
\cite{MR,FNTW,MIR}. More recently \cite{FNS, FZ} it has been
argued that the non-Abelian statistics might come out from
constraining an Abelian theory, by employing the Meissner effect
in the neutral sector; that would also account for the pairing
phenomenon.

Lately a Conformal Field Theory (CFT) description in terms of
composite fermions for the Jain filling fractions
$\nu=\frac{m}{2pm+1}$ of a Quantum Hall Fluid (QHF) has been
proposed has been proposed \cite{CGM, PS}. The composite fermions
are described by composite vertex operators, which are (the image
of) primary fields of a CFT with central charge $c=m$ in the
Lowest Landau Level (LLL), and factorize into a charged and a
neutral component. The neutral component assures the locality
properties of the composite electron field and the
single-valuedness of the ground state wave function.

Further due to the $Z_m$ symmetry in the neutral sector the
general $N_e$ electron correlation function shows clustering
properties which have been proposed in the context of paired Hall
states \cite{RR}.

In this letter we present a natural extension of the
$m$-reduction procedure (previously applied in ref.\cite{CGM} to
the description of the Jain fillings) to the case of the
non-standard fillings $\nu=\frac{m}{pm+2}$. In particular for
$m=2$ the neutral degrees of freedom of the composite electrons
describe Majorana fermions and contribute to the ground state
wave function with a Pfaffian term, in agreement with an early
proposal \cite{MR}. Further, for generic $m$, the neutral degrees
of freedom are parafermions and give rise to the clustering
phenomenon of $m$ objects, which simply reproduces the picture
presented in ref.\cite{RR} by employing a hamiltonian formalism
with $m+1$ body interactions.

The letter is organized as follows: after a brief review of the
$m$-reduction procedure we construct the parafermion vertex
operators which together with the charged $U(1)$ component give
rise to the primary fields of the CFT. It should be noticed an
interesting phenomenon related to the existence of two classes of
primary fields: the $Z_m$ twist invariant and the $Z_m$
non-invariant ones. In fact on physical grounds one can argue
that only the invariant fields are relevant to the description of
the pairing phenomenon. As a result the central charge of neutral
sector of our CFT is given by $c=\frac{2(m-1)}{m+2}$. An
interpretation of such a phenomenon has been proposed in
ref.\cite{CT} in terms of a strong coupling between the symmetric
modes.

We then give the ground state wave function as a correlator of
the primary fields with integer electric charge and the
quasi-holes wave function for the elementary excitations, with
fractional charge, up to the correlator of the twist fields.

Our approach is meant to describe all the plateaus with even
denominator starting from the bosonic Laughlin filling $\nu=1/2$,
which is described by a CFT with $c=1$, in terms of a scalar
chiral field compactified on a circle with radius $R^2=1/\nu=2$
(or the dual $R^2=1/2$). Then the $U(1)$ current is given by
$J(z|1,2)=i\partial_z Q(z)$, where $Q(z)$ is the compactified
Fubini field with the standard mode expansion:
\be
Q(z)=q-i\, p\, ln z + \sum_{n\neq 0}\frac{a_n}{n}z^{-n}
\ee
with $a_n$, $q$ and $p$ satisfying the commutation relations $
\left[a_n,a_{n'}\right]=n\delta_{n,n'}$ and $\left[q,p\right]=i $.

The representations are realized by the vertex operators $
U^{\alpha}(z)=:e^{i\alpha Q(z)}: $ with $\alpha^2=2$ and
conformal dimension $h=1$, with a consequent extension of the
$U(1)$ symmetry to the $SU(2)_1$ affine one. Furthermore the
theory contains the Virasoro algebra generated by the
stress-energy tensor $T(z|1,2)=-\frac{1}{2}:\left(\partial_z
Q(z)\right)^2:$.

In order to construct the $\nu=m/2$ filling we start with the set
of fields in the above CFT (mother). Using the $m$-reduction
procedure, which consists in  considering the subalgebra
generated only by the modes which are divided by an integer $m$,
we get the image of an orbifold of a $c=m$ CFT (see
ref.\cite{CGM} and references therein).

Also for the $SU(2)$ case the fields in the mother CFT can be
factorized into irreducible orbits of the discrete group $Z_m$
which is a symmetry of the daughter theory and can be organized
into  components which have well defined transformation
properties under this group.

In order to compare the image so obtained to the $c=m$ CFT, we map
$z\rightarrow z^{1/m}$ and we will indicate the components in the base
$\hat{z}=z^{m}$ with an hatted symbol (for instance, $\phi(z)\rightarrow
\widehat{\phi}(z)$). In particular any component in the subalgebra is a
function only of the variable $z^m$.

In ref.\cite{VM} it was also defined an isomorphism between fields on the
$z$ complex plane and fields on the $z^m$ plane by means of the following
identifications:
\be
a_{nm+l} \longrightarrow \sqrt{m}a_{n+l/m} \hspace{1cm} q
\longrightarrow \frac{1}{\sqrt{m}}q  \label {eq: 32}
\ee

Let us first introduce the invariant scalar field
\be
X(z|m,2)=\frac{1}{m}\sum^{m}_{j=1}Q(\varepsilon^j z)
\ee
where $\varepsilon^j=e^{i\frac{2\pi j}{m}} $, corresponding to a
compactified boson on a circle with radius now equal to $R_X^2=
2/m$. This field depends only on powers of $z^m$ and satisfies
trivial boundary conditions. It is the basic field of the $U(1)$
electrically charged sector of the theory where the charge is
measured by the zero mode.

The non-invariant components, as a resulting image of $m$
constrained bosons, are expressed by
\be
\phi^j(z|m,2)=Q(\varepsilon^j z)-X(z|m,2)
\ee
with the condition $\sum_{j=1}^{m}\phi^j(z|m,2)=0$.

These fields satisfy non-trivial twisted boundary conditions
\be
\alpha{\cdot}\phi^j(\varepsilon z|m,2)=\alpha{\cdot}\phi^{j+1}(z|m,2)+2\pi n \alpha{\cdot}p
\,\,\, n\in Z \label{eq: shift}
\ee
where the shift is due to the definition of index $j$ mod $m$.

The $J(z|1,2)$ current of the mother theory decomposes into a
charged current given by $J(z|m,2)=i\partial_z X(z|m,2)$ and
$m-1$ neutral ones $\partial_z\phi^j(z|m,2)$.

In the same way every vertex operator in the  mother theory can
be factorized in a vertex that depends only on the invariant
field:
\be
{\cal U}^{\alpha}(z|m,2)=z^{\frac{\alpha^2}{2}\frac{(m-1)}{m}}:e^{i\alpha{\cdot}
X(z|m,2)}: ~~~\alpha^2=2
\ee
and in vertex operators depending on the $\phi^j(z|m,2)$ fields.

We also introduce the neutral components:
\be
\psi^{\alpha}_1(z|m,2)=\frac{z^{\frac{\alpha^2}{2}\frac{(1-m)}{m}}}{m}
\sum_{j=1}^{m}
\varepsilon^{\frac{\alpha^2 j}{2}} :e^{i\alpha{\cdot} \phi^{j}(z|m,2)}:
\ee
which satisfy the fundamental product:
\be
\psi_1^{\alpha}(z|m,2)\psi_1^{\beta}(\xi |m,2)
\nn \\ =
\frac{z^{\frac{\alpha^2}{2}\frac{1-m}{m}}
\xi^{\frac{\beta^2}{2}\frac{1-m}{m}}}{m^2}
\sum_{j,j'=1}^{m}\varepsilon^{\frac{\alpha^2 j+\beta^2 j'}{2}}
:e^{i\alpha{\cdot}\phi^{j'}(z|m,2)}
e^{i\beta{\cdot}\phi^j(\xi |m,2)}:
\frac{(\varepsilon^{j'}z-\varepsilon^{j}\xi)^{\alpha
{\cdot} \beta}}{(z^m-\xi^m)^ {\frac{\alpha {\cdot} \beta}{m}}}
\label {eq: 31}
\ee

The set of primary fields generated by this product can be
expressed in terms of the fundamental representations $\Lambda^i$
of $SU(m)$ Lie algebra. In fact, defining
\be
\phi^{\Lambda^i}(z|m,2)=\sum_{j=1}^{i}\phi^j(z|m,2)
\ee
and $\phi^{\Lambda}(z|m,2)$, where
$\Lambda=\sum_{i=1}^{m-1}l_i\Lambda_i$, and introducing the
$m$-ality parameter $a=\sum_{i=1}^{m-1}i l_i$ (mod $m$), which is
invariant under the addition of any vector in the root lattice,
the exact form of these fields can be deduced by the analysis of
the OPE of eq.(\ref{eq: 31}) for $\alpha=\beta$ to get the $a=2$
field. By repeated application of this analysis we can obtain the
full set of fields:
\be
\widehat{\psi}^{\alpha}_a(z|m,2)=
\sum_{j_1>j_2>\dots >j_a =1}^{m}f(\varepsilon^{j_1},\dots,\varepsilon^{j_a},z^{1/m})
:e^{i\alpha\widehat{\phi}^{j_1}(z|m,2)}\dots e^{i\alpha\widehat{\phi}^{j_k}(z|m,2)}:
\ee
where the functions $f(\varepsilon^{j_1},\dots,\varepsilon^{j_a},z)$ can be
extracted from the OPE relations.

The sum takes into account the fact that any field can be associated to the
$a$-th fundamental representation of $SU(m)$ (namely, the antisymmetric
tensor representation). In \cite{VM} it was shown that these are a
realization of parafermions and satisfy the operator product algebra
\cite{GQ}:
\be
\widehat{\psi}_{a} ^{\alpha}(z|m,2) \widehat{\psi}_{a'} ^{\alpha}(\xi |m,2)
=\frac{C_{a,a'}}{(z-\xi)^{\frac{2 a a'}{m}}}
\left[\widehat{\psi}_{a+a'} ^{\alpha}(z|m,2)+O(z-\xi)\right] \,\,\, a+a'<m
\ee
and for $a+a'=m$ the OPE contains the Virasoro algebra
generators:
\be
\widehat{\psi}_{a} ^{\alpha}(z|m,2) \widehat{\psi}_{m-a} ^{\alpha}(\xi |m,2)
=\frac{C_{a,m-a}}{(z-\xi)^{\frac{2 a (m-a)}{m}}}
\left[1+ \frac{2 h_a}{c_\psi}\widehat{T}_{\psi}(z|m,2)(z-\xi)^2+O(z-\xi)^3\right]
\ee
where $C_{a,a'}$ are the structure constants and  $h_a$, $c_\psi$
are the conformal dimensions and central charge of parafermions
\cite{GQ}.

Moreover the $SU(m)$ representations that can appear are the
fundamental ones $\Lambda_a$, because in the OPE algebra of
$\psi^{\alpha}_1(z|m,2)$ (corresponding to the $\Lambda_1$
representation) only the fields $\psi^{\alpha}_{a}(z|m,2)$ with
$a\in \{1,\dots , m-1\}$ appear while $\psi^{\alpha}_{m}(z|m,2)$
is the identity operator. That is an effect of the $Z_m$
invariance of the parafermions algebra.

Notice that no neutral currents are present in the above OPE, as one would
expect from symmetry considerations.

It is well known that the $c_X=1$ {\em rational}CFT with
$R_X^2=2/m$ has $2m$ primary fields which can be parametrized by
$\alpha=\sqrt{\frac{2}{m}}a$ and $\alpha=\sqrt{\frac{1}{2m}}a$,
$~~ a=1,\dots ,m$. In our formalism these fields appear together
with the neutral ones into the composite operators:
\be
\widehat{V}^{\sqrt{\frac{2}{m}}a}(z|m,2)=\widehat{{\cal
U}}^{\sqrt{\frac{2}{m}}a}(z|m,2)\widehat{\psi}_a(z|m,2)
\label {eq: ANI}
\ee
for quasi-particles and
\be
\widehat{V}^{\frac{a}{\sqrt{2 m}}}_{qh} (z|m, 2)=
\widehat{{\cal U}}^{\frac{a}{\sqrt{2 m}}} (z|m, 2) \hat{\sigma}_a (z|m,2)
\label {eq: QH}
\ee
for quasi-holes; $\hat{\sigma}_a(z|m, 2)$ are the parafermionic
twist fields. The electric charges and magnetic flux contents of
such fields is given after eqs.(\ref{eq: CQP}, \ref{eq: CQH}).

In eq.(\ref{eq: QH}) we have not included the contribution coming
from the $\bar{\sigma}$ fields of the $Z_m$ non-invariant theory.
In fact they give rise to a term which does not survive after the
projection to the LLL\footnote{We notice that we do not have an
explicit realization of the $\bar{\sigma}$ fields because we have
projected the mother $c=1$ CFT onto the $m$-covering of the
plane, so that those fields appear as branch cuts \cite{ZK,
VM4}.}.

The currents $\widehat{V}^{{\pm}\sqrt{\frac{2}{m}}}(z|m, 2)$ and
$J(z|m, 2)$ generate the $SU(2)_m$ affine algebra while
$\widehat{V}^{\frac{a}{\sqrt{2 m}}}_{qh}$ are the primary fields
of this algebra (see ref.\cite{GQ} for details).

The generator of the Virasoro algebra $\hat{T}(z|m,2)$ was given in
\cite{VM2} as the sum of two independent operators, one depending on the
charged sector:
\be
\widehat{T}_{X}(z|m,2)=-\frac{1}{2} :\left(\widehat{\partial_{z}X}(z|m,2)\right)^2:
\ee
and the other given in terms of the $Z_m$ twisted bosons
$\widehat{\phi}^j(z|m,2)$:
\be
\widehat{T}_{\psi}(z|m,2)=\frac{2}{m+2}
\left(-\sum_{j=1}^{m}\frac{:(\widehat{\partial_{z}\phi}^j(z))^2:}{2 m^2}+
\sum_{j'\neq j=1}^{m}\frac{\varepsilon^{j'+j}
:e^{-i \alpha \widehat{\phi}^{j'}(z)}e^{i \alpha \widehat{\phi}^j(z)}:}{
2 m^2 z^2 (\varepsilon^{j'}-\varepsilon^{j})^2}+\frac{m^2-1}{24 m
z^2}\right) \label {eq: STRESS}
\ee
while the higher integer spin operators generating the full
parafermionic ${\cal W}_m$ algebra \cite{FZL} was given in terms
of the $\widehat{\phi}^j(z|m,2)$ fields in \cite{VM3}. Notice
that the vacuum expectation value of $\widehat{T}_{\psi}$ is zero
due to the cancellation between the second and the third term in
eq.(\ref {eq: STRESS}).

It is not very hard to verify that the conformal dimensions of the fields
of eq.(\ref{eq: ANI}) and eq.(\ref{eq: QH}) are given by:
\be
h_a=\frac{a^2}{m}+ a \left(\frac{m-a}{m}\right) = a
\,\,\,\, a\in\{1,\dots,m\}
\ee
for quasi-particles and
\be
h^{qh}_a=\frac{a^2}{4 m}+\frac{a(m-a)}{2 m (m+2)} = \frac{a(a+2)}{4(m+2)}
\,\,\,\, a\in\{1,\dots,m\}
\ee
for quasi-holes.

The contribution to the central charge $c$ is given by
\be
c=1+\frac{2(m-1)}{m+2} = \frac{3 m}{m+2}
\ee
where 1 and ${\displaystyle \frac{2(m-1)}{m+2}}$ come from the charged and
neutral degrees of a freedom respectively.

We must notice that the contribution of the $Z_m$ invariant
fields to the central charge is not equal to $m$. In fact the
missing part is associated with the non-invariant fields defined
as:
\be
\psi^{\alpha(i,j)}(z|m,2)=\frac{z^{\frac{\alpha^2}{2}\frac{(1-m)}{m}}}{m}\left(
\varepsilon^{\frac{\alpha^2 i}{2}} :e^{i\alpha{\cdot} \phi^{i}(z|m,2)}:-
\varepsilon^{\frac{\alpha^2 j}{2}} :e^{i\alpha{\cdot} \phi^{j}(z|m,2)}:\right)
\ee
with $i\neq j =\{1,\dots ,m\}$, which are primary fields of the
complementary $\bar{c}=\frac{m(m-1)}{m+2}$ theory. One can see
also that these extra degrees of freedom can be gauged away by
means of a coset reduction.

This appears to be a generalization of the Low-Barrier limit of a
double-layer sample described in ref.\cite{CT} where a strong
tunnel effect was introduced. In our case the strong coupling
between electrons of the different layers is simply a consequence
of the induced $Z_m$ symmetry of the composite electrons states.
This phenomenon was suggested to be relevant for CFT's on
algebraic curves in \cite{VM4}. It would be interesting to have a
clear picture of the symmetries of the vacuum and excited states,
so to understand the mechanism which decouples the $Z_m$
non-invariant theory from the invariant one.

We can now describe the generic filling $\nu = {\displaystyle
\frac{m}{p m +2}}$ by flux attachment starting from $\nu = m/2$.
In order to do so, we factorize the fields into two parts, the
first describing the $c_X=1$ charged sector with radius
$R_X^2=\frac{pm+2}{m}$, the second describing the neutral
excitations, which are parafermions, with central charge
$c_{\psi}=\frac{2(m-1)}{m+2}$, for any $p\in N$.

The $U(1)$ sector is now described by the compactified boson $X(z|m,pm+2)$
and its related vertex operators ${\cal U}^{{\pm}\alpha_l}(z|m,pm+2)$, with
$\alpha_l=l/\sqrt{m(pm+2)}$, $l=1,\dots , m (pm+2)$, which produce
excitations with anyonic statistics $\theta=\pi\alpha_l^2$. While the $m-1$
neutral bosons $\phi^j(z|m,2)$ are independent from the flux number $p$.

To obtain a pure holomorphic function we will consider the
correlator of the composite operators
$\widehat{V}^{\alpha_l}(z|m,pm+2)$ with conformal dimensions:
\be
h_l=\frac{l^2}{2m(pm+2)}+a\left(\frac{m-a}{m}\right) ~~~l=(pm+2)a
~~~a=1,2,\dots ,m \label {eq: CQP}
\ee
They describe dressed $a$-electrons with electric charge
$q^e_a=a$ and ``magnetic charge" $q^m_a=\frac{pm+2}{m}a$
interacting through the neutral ``cloud" associated with them
(see eq.(\ref{eq: ANI})).

Also we will consider correlators in which are present quasi-hole operators
given by $\widehat{V}_{qh}^{\alpha_l}(z|m,pm+2)$ having conformal
dimensions:
\be
h^{qh}_l=\frac{l^2}{2m(pm+2)}+\frac{a(m-a)}{2m(m+2)}
\label {eq: CQH}
\ee
where $(pm+2)(a-1)<l<(pm+2)a$ and $a=1,2,\dots ,m$.

Their electric and magnetic charges are $q^e_l=\frac{l}{pm+2}$
and $q^m_l=\frac{l}{m}$.

We should point out that $m$-ality in the neutral sector is
coupled to the charged one in analogy to the case of Jain
hierarchical fillings in order to assure the locality of
electrons with respect to all the edge excitations \cite{CGM}.
This follows from the fact that our projection when applied to a
local field automatically couples the discrete $Z_m$ charge of
$U(1)$ with the neutral sector, in order to give a totally
single-valued composite field.

Also notice that the $m$-electron vertex operator does not
contain any neutral field. Therefore, the $m$-electron wave
function is realized only by means of the $c_X=1$ charged sector
as proposed in ref.\cite{CMNN}.

We are now ready to give the holomorphic part of the ground state wave
function for the generic filling $\nu=\frac{m}{pm+2}$. To such an extent we
consider the $N_e$ single($a=1$)-electrons correlator which factorizes into
a Laughlin-Jastrow type term coming from the charged sector:
\be
<N_e\alpha|\prod_{i=1}^{N_e}\widehat{{\cal
U}}^{\alpha}(z_i|m,2pm+2)|0>
=\prod_{i<i'=1}^{N_e}(z_i-z_{i'})^{p+\frac{2}{m}}
\ee
and a contribution coming from the neutral excitations:
\be
<0|\prod_{i=1}^{N_e}\widehat{\psi}_1
^{\alpha}(z_i|m,2)|0>
=\frac{\sum_{\{j_{i}\}=1}^{m}\varepsilon^{\frac{\alpha^2}{2}(2i-1) j_i+ j_{i}}
\prod_{\{j_{i},j_{i'}\}=1}^{m}
(\varepsilon^{j_{i}}z^{1/m}_i-\varepsilon^{j_{i'}}
z^{1/m}_{i'})^{\alpha^2}}{\prod_{i<i'=1}^{N_e}(z_{i}-z_{i'})^{\frac{2}{m}}}
\label{eq: FUN}
\ee

For $N_e$ multiple of $m$ we observe that the non analytic part
of the neutral fields $\widehat{\psi}^{\alpha}(z_i|m,2)$ is
necessary to eliminate the non integer part of the exponent in
the correlator of the charged fields. We also point out that in
our formalism the correlators are given, in principle, for any
$m$ and $N_e$ and that follows from the projection procedure
only.

By considering the case $m=2$, $p$ odd that is for $\nu=1/q$, $q=p+1$ we
get for the correlator of $N_e$ single($a=1$)-electrons:
\be
<N_e\alpha|\prod_{i=1}^{N_e}\widehat{V}^{\sqrt{2q}}(z_i|2,2q)|0>
=\prod_{i<i'=1}^{N_e}(z_i-z_{i'})^{q}Pf\left(\frac{1}{z_i-z_{i'}}\right)
\ee
(where $Pf\left(\frac{1}{z_i-z_{i'}}\right)={\cal
A}\left(\frac{1}{z_1-z_2}\frac{1}{z_3-z_4}\dots\right)$ is the
antisymmetrized product over pairs of electrons) which is in
agreement with previous results \cite{MR,RR}. For the generic $m$
case, even though it is hard to work out explicitly the sum over
the phases in eq.(\ref{eq: FUN}), for the $N_e$ point functions
we found the clustering properties previously given in
ref.\cite{RR} in another framework by grouping particles into
clusters of $m$.

In this last case the neutral modes describe parafermions and
contribute to the ground state wave function with a generalized
Pfaffian term. It would be very interesting to give an
interpretation of the Pfaffian term (but also of its
generalization presented in eq.(\ref {eq: FUN})) in the context
of a plasma description a la Laughlin in order to better
understand the physics of the paired states \cite{CHMMZ}.

In a similar way we also are able to evaluate correlators of $N_e$
single($a=1$)-electrons in the presence of quasi-hole excitations. In
particular for $m=2$ and for two quasi-holes we get:
\bea
&&\frac{<N_e\sqrt{2q}+2/\sqrt{2q}|\prod_{i=1}^{N_e}\widehat{V}^{\sqrt{2q}}(z_i|2,2q)
\widehat{V}^{1/\sqrt{2q}}(w_1|2,2q)\widehat{V}^{1/\sqrt{2q}}(w_2|2,2q)|0>}
{<2/\sqrt{2q}|\widehat{V}^{1/\sqrt{2q}}(w_1|2,2q)\widehat{V}^{1/\sqrt{2q}}(w_2|2,2q)|0>}
\nn  \\
&&=\prod_{i<i'=1}^{N_e}(z_i-z_{i'})^{q}
Pf\left(\frac{(z_i-w_1)(z_{i'}-w_2)+(z_{i'}-w_1)(z_{i}-w_2)}{z_i-z_{i'}}\right)
\label {eq: TWS}
\eea
in agreement with the wave functions proposed in Ref.\cite{MR}.

The explicit evaluation of correlation functions for the twist
fields lies outside the scope of the present paper, but it does
not affect the main results given in eq.(\ref {eq: TWS}). On the
other hand, it is fundamental to understand the non-Abelian
statistics of the quasi-holes. We will analyze this aspect in a
forthcoming paper.

We also point out that we are not considering the full set of
primary fields in the theory. Indeed, also for $p=0$ there are
neutral fields which correspond to the ``termal" fields of
parafermionic theory \cite{GQ}.

\bigskip
{\bf Acknowledgments} - We thank M. Huerta, G. Zemba  and A.
Sciarrino for useful comments and for reading the manuscript.

\end{document}